\documentclass[referee]{aa}
\usepackage{graphicx}
\usepackage{txfonts}
\begin{document}


\title{Evidence for collisional depolarization of the   \mbox{\ion{Ba}{ii}  ${\lambda}4554$}   line in the low chromosphere}
\author{M. Derouich\thanks{Also associated researcher at CNRS UMR 8112 -- LERMA,   
Observatoire de Paris, Section de Meudon, F-92195 Meudon, France}}
\institute{Institut d'Astrophysique Spatiale, CNRS-Universit\'e Paris-Sud 11, 91405 Orsay Cedex, France
}
\titlerunning{Scattering polarization of the \ion{Ba}{ii}  ${\lambda}4554$   line}
\authorrunning{M. Derouich}

\date{  Accepted for publication in Astronomy and Astrophysics}

\abstract
{Rigorous modeling of the   \ion{Ba}{ii} ${\lambda}4554$ formation  is potentially interesting  since  this strongly polarized line  forms  in the solar chromosphere where the magnetic field  is rather poorly known.
} 
{ To investigate the role of    isotropic collisions with neutral hydrogen in the   formation of the polarized   \ion{Ba}{ii} ${\lambda}4554$  line and, thus, in the determination of the  magnetic field. 
}
{
Multipole relaxation and transfer rates of the    $d$ and  $p$-states of  \ion{Ba}{ii}  by  isotropic collisions with neutral hydrogen are calculated.
 We consider a plane parallel layer of  \ion{Ba}{ii} situated at the low chromosphere and anisotropically illuminated  from below which produces   linear polarization in the    ${\lambda}4554$ line   by scattering processes. To compute that polarization, we   solve the statistical equilibrium equations for   \ion{Ba}{ii}  levels including collisions, radiation and magnetic field effects.
}
{Variation laws of the relaxation and transfer rates  with  hydrogen number density $n_{\textrm {\scriptsize H}}$  and temperature are deduced.    The   polarization  of the ${\lambda}4554$  line is  clearly affected  due to isotropic collisions with neutral hydrogen  although the collisional depolarization  of its upper level   $^2P_{3/2}$    is negligible. This is because the alignment  of the  metastable  levels  $^2D_{3/2}$ and $^2D_{5/2}$ of the \ion{Ba}{ii} are vulnerable to collisions. At   the height of
formation of the ${\lambda}4554$  line where  $n_{\textrm {\scriptsize H}} \sim 2 \times  10^{14}$ cm$^{-3}$, we find that the neglecting of the   collisions  induces   inaccuracy of $\sim$ 25\% on the calculation of the polarization and   $\sim$ 35  \%   inaccuracy on     microturbulent   magnetic field determination.
}
{The polarization of the  ${\lambda}4554$  line decreases  due to collisions  with hydrogen atoms.  In addition, during scattering processes collisions  could change the frequency of the \ion{Ba}{ii} photons.  To quantitatively study this line, one should confront   the problem of development of general theory treating partial redistribution of frequencies and including   
 transfer and relaxation  rates     by collisions for a multilevel atom with hyperfine  structure. 
}
 
\maketitle
\section{Introduction}
Depolarization due to the Hanle effect can be  used  to retrieve information on solar magnetic fields  (e.g. Trujillo Bueno    2001). This magnetic depolarization is   usually  mixed with a collisional depolarization. To quantitatively  use the Hanle effect as a technique of investigation of solar magnetic fields,  one needs to know  the collisional rates.  In addition, one    should   include properly both the collisional  depolarizing and  {\it transfer}  rates in the line formation modeling. For instance,   Derouich et al. (2007)   found that the effect of the collisions, in   typical solar conditions of temperature and hydrogen density,  is particularly important for the very important  Ti {\sc i}  ${\lambda}4536$ line--  a collisional depolarization of   more than 25 \% is obtained. In fact, at a first glance, because the inverse lifetime of the  upper level is larger than  the value of the elastic depolarizing rate  $D^2$, one might think (wrongly) that the effect of the collisions is negligible for the Ti {\sc i}  ${\lambda}4536$ line.   Collisional  depolarization  of this line  is mainly due to  {\it transfer rates}   which are generally neglected (see Eq. 14 of Derouich et al.  2007).   

Since their ionization potential is rather low, most Barium atoms are ionized throughout the
  low chromosphere  (Tandberg-Hanssen \& Smythe 1970).
 The core of the Ba {\sc ii} ${\lambda}4554$ line is formed at about  800 km
above the photosphere (e.g.  Uitenbroek \& Bruls  1992). The wings are formed in deeper layers of the photosphere. Recently, a renewed interest in the measurement and the physical interpretation  of the linear polarization  of the   Ba {\sc ii} ${\lambda}4554$ line have emerged (e.g. Malherbe et al., 2007; Belluzzi et al. 2007; Lopez Ariste et al. 2008).  It has been shown that during the formation   of this line the hyperfine structure and the partial redistribution of the frequencies effects are  very important.  
 In this work,  we   investigate fully the   importance of isotropic collisions with neutral hydrogen in its   modeling (Sects. 2 and 3).   Sect. 4 is dedicated to the calculation of the impact of    neglecting collisional effects on the magnetic field determination. Our concluding remarks are given in Sect. 5. 
      
  \section{Modeling of the creation of the polarization in the Ba {\sc ii} levels}
Let us consider a plane-parallel layer in the low chromosphere ($\sim$ 800 km above the photosphere) formed by Ba {\sc ii} ions and  illuminated   anisotropically  by  the photospheric continuum  
 radiation  field. In a weakly polarizing medium like the solar atmosphere one can safely neglect the contribution of polarization to the excitation of Ba {\sc ii} ions. Assuming   that the incident radiation has cylindrical symmetry around the local solar vertical through the scattering center, only the multipole orders $k$ = 0 (mean intensity $J^0_0$) and $k$ = 2 (radiation tensor $J^2_0$ associated to the anisotropy) are needed to fully describe the incident radiation.  For each wavelength, the value of the anisotropy factor ($w=\sqrt{2}J^2_0/J^0_0$) and of the number of photons per mode ($\bar{n}=J^0_0(c^2/2h\nu^3)$) are obtained from   Fig. 2 of Manso Sainz \& Landi Degl'Innocenti (2002).  In particular, for  the Ba {\sc ii} ${\lambda}4554$ line, we obtain $w=0.16$ and $\bar{n}=2.59   \times 10^{-3}$ which are similar to the values calculated by Belluzzi et al. (2007).  We verified that a reasonable modification of $w$ and  $\bar{n}$ (i.e.  by around 10\%) do not affect  the results of the present  work. Obviously, a  realistic simulation of the  line formation conditions requires a careful  consideration of the transfer of radiation in a medium that is not optically thin. This problem   is out of the scope of this paper.  The   formulae  we  use   to compute the linear polarization    for the case of a tangential observation in a plane-parallel atmosphere   is   presented, for example, in Sect. 2  of Derouich et al. (2007).

\begin{figure}
\begin{center}
\includegraphics[width=6cm]{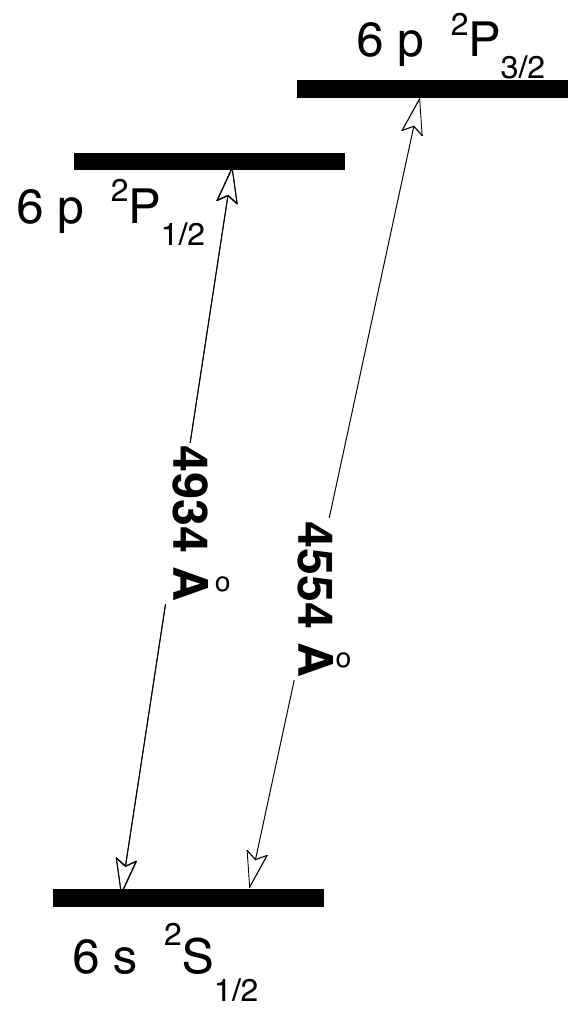}   
\end{center}
\caption[]{Partial Grotrian diagram of  Ba {\sc ii} showing the levels and the spectral wavelengths taken into account in the case of  the simplified model. Note that the level spacings are not to scale.}
\label{figure1}
\end{figure}

\begin{figure}
\begin{center}
\includegraphics[width=8cm]{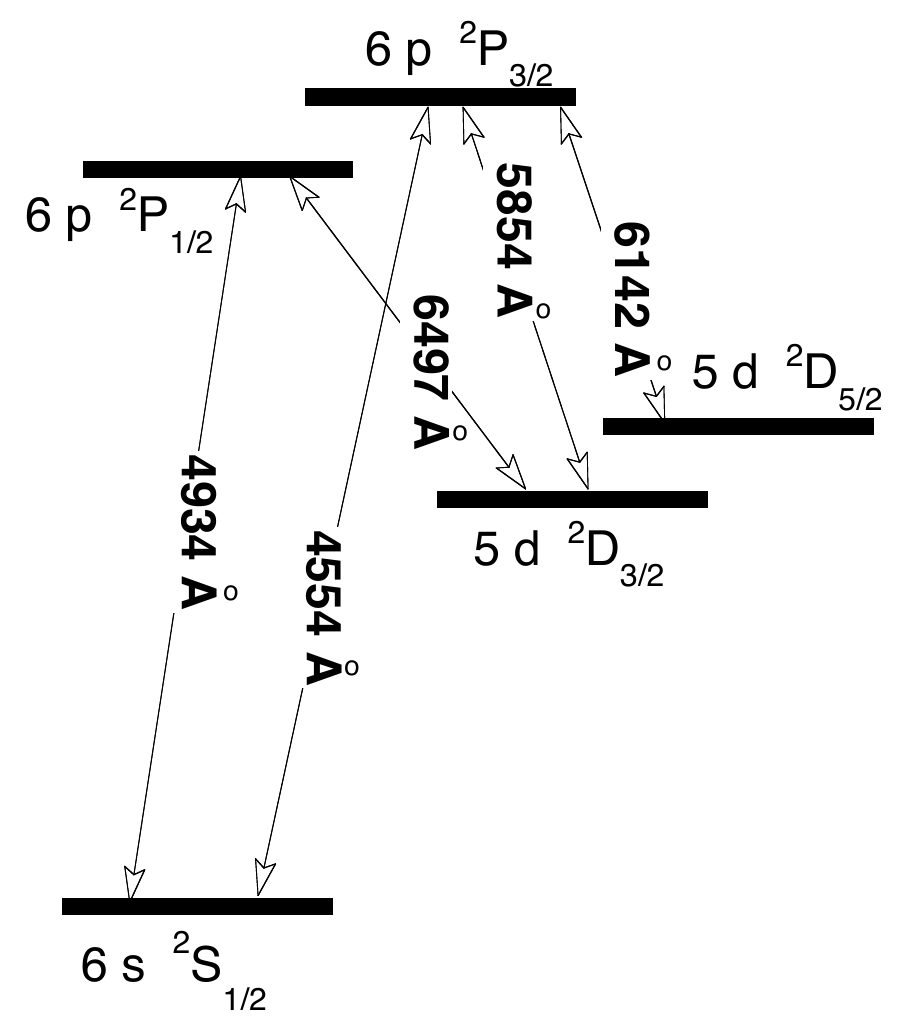}   
\end{center}
\caption[]{Partial Grotrian diagram of  Ba {\sc ii} showing the levels and the spectral lines of the 5 levels-5 lines model. Note that the level spacings are not to scale.}
\label{figure2}
\end{figure}

 A simplified  atomic model approximation (Fig. \ref{figure1})    can reduce  considerably the numerical and theoretical calculations especially if the  transfer of radiation  effects are   considered.   However, the hypothesis of neglecting   metastable $d$-sates in the  Ba {\sc ii} ${\lambda}4554$ line modeling    is highly questionable. We   calculate the polarization $p=Q/I$ using the   simplified atomic model (see Fig. \ref{figure1}) and the 5 levels-5 lines  model   (see Fig. \ref{figure2}) which accounts for the metastable $d$-levels. We find that the simplified model  overestimates the polarization degree of the Ba {\sc ii} ${\lambda}4554$:  
 \begin{eqnarray}  
\frac{ {(p)}_\textrm{5 levels-5 lines model} } { {(p)}_\textrm{simplified model} } \simeq 0.7 
\end{eqnarray}   
Thus, a more realistic diagnostic of the polarization of this line should be performed in the framework of 5 levels-5 lines model.    In the description of the Ba {\sc ii}, we neglect  the
contribution of the hyperfine structure.  There are no additional conceptual difficulties for  including the 
contribution of the hyperfine relaxation and polarization transfer rates by isotropic collision.  Our main conclusions concerning the evidence for collisional depolarization of the    \ion{Ba}{ii}  ${\lambda}4554$    line  remain unchanged.   

   \section{Contribution of collisions in the framework of the 5 levels-5 lines model}  
\subsection{Definitions of the depolarization, relaxation, and polarization transfer rates and statistical equilibrium equations}
The collisional  evolution of the density matrix  components $\rho_{q}^{k} (J)$   is due   to the     gain-terms  denoted as polarization transfer rates  and to  the loss-terms denoted as relaxation rates.      The polarization transfer rates, denoted by  $C_I(k   \;  J  \leftarrow  k   \;  J', \; T)$  if $E_J  >  E_J'$  and  $C_S(k   \; J   \leftarrow  k    \; J')$ if $E_J  <  E_J'$, are  due to inelastic and super-elastic collisions respectively. $E_J$ being the energy of the $J$-level.   The left arrow ($\leftarrow$) is used because these  rates   represent gain-terms    in the  statistical equilibrium equations  (SEE) describing the evolution of the $J$-level.  

The relaxation  rates are the sum  of the term $D^k(J, \; T)$ exclusively responsible  for the depolarization of the level $(J)$ and the terms  $C_I(k=0 \;  J \to  k=0  \; J', \; T)$ ($E_J  <  E_J'$)  and $C_S(k=0 \;  J \to  k=0  \; J', \; T)$ ($E_J >  E_J'$)  corresponding to the population transfer between the levels $J$ and $J'$.  We  notice that,  in principle, the denotations ``inelastic'' and ``superelastic''   are used for transitions between different electronic sates. Within  the same spectral term (inside a given electronic state), collisions are  called ``elastic'' even      these involving transitions between two different $J$-levels.  However, we adopt the  denotations $C_S$ and $C_I$ since they are usually used in astrophysics\footnote{Apart a multiplicity factor of $\sqrt{ \frac{2J'+1} {2J+1}}$, we use the same definitions of the monograph  by Landi Degl'Innocenti \& Landolfi (2004).}. We denote by $R^k (J, \; T)$   the relaxation rates of rank $k$ given by:
\begin{eqnarray}  
R^k (J, \; T) = D^k(J, \; T)+ R^0 (J, \; T)= D^k(J, \; T)&+&\sum_{J_l}    \sqrt{ \frac{2J_l+1} {2J+1}}  \;  C_S(k=0 \;  J   \to k=0 \; J_l, \; T)  \\ && + \sum_{J_u}    \sqrt{ \frac{2J_u+1} {2J+1}} \; C_I(k=0 \;  J   \to k=0 \; J_u, \; T) \nonumber
\end{eqnarray}
The overall population  of the level $J$ is not affected by purely elastic collisions, $D^0 (J, \; T)$=0.
  The  SEE are:  
  \begin{eqnarray} \label{SEE}
\big[\frac{d}{dt}\; \rho_q^{k} (J)\big]_{\rm coll} & = &  
  - R^{(k)}(J, \; T)   \; \rho_q^{k} (J)  + \sum_{J_l}  
C_I(k  \; J  \leftarrow k  \; J_l, \; T) \;  \rho_q^{k} (J_l)        \nonumber \\  &&
+ \sum_{J_u } C_S(k  \; J \leftarrow  k  \; J_u, \; T) \;  \rho_q^{k}(J_u)  
\end{eqnarray} 
 Despite
 $-k \; \le \; q \; \le \;k$,  due to the isotropy of the collisions with neutral hydrogen, the relaxation and polarization transfer rates are calculated only for $ q =0   $ since they are  $q$-independent. The effect of the isotropic collisions is the same for   density matrix elements  with  $ q =0   $   and for off-diagonal  ones corresponding to coherences  where $ q \ne 0$.   

 In  the standard case where polarization phenomena are neglected,  the SEE can be obtained from Eq. (\ref{SEE}) simply by retaining the terms with $k=q=0$:
\begin{eqnarray}   \label{SEEstandard}
\big[\frac{d}{dt}\; \rho_0^{0} (J)\big]_{\rm coll} & = &  
  -  \big[\sum_{J_l}    \sqrt{ \frac{2J_l+1} {2J+1}}  \;  C_S(k=0 \;  J   \to k=0 \; J_l, \; T) \nonumber \\ && + \sum_{J_u}    \sqrt{ \frac{2J_u+1} {2J+1}} \; C_I(k=0 \;  J   \to k=0 \; J_u, \; T)\big]  \; \rho_0^{0} (J)   \\  && + \sum_{J_l}  
C_I(k=0  \; J  \leftarrow k=0  \; J_l, \; T) \;  \rho_0^{0} (J_l)       
+ \sum_{J_u } C_S(k=0  \; J \leftarrow  k=0  \; J_u, \; T) \;  \rho_0^{0}(J_u)  \nonumber
\end{eqnarray} 
 Since,
\begin{eqnarray} \label{eq_10}
\sqrt{ \frac{2J'+1} {2J+1}}  \; C (k=0 \;  J   \to k=0 \; J', \; T) &=& \zeta(J \to J')   \nonumber \\
 \sqrt{2J+1} \;  \rho_0^{0} (J) &=& N_J 
\end{eqnarray} 
where $\zeta(J \to J') $ is the usual usual fine structure collisional transfer rate and $N_J$ is the population of the level $J$, Eq. (\ref{SEEstandard})  becomes (with evident notations):
\begin{eqnarray} 
\big[\frac{d}{dt}\; N_J\big]_{\rm coll} & = &  
  -  \big[\sum_{J_l}      \;  \zeta_S (J   \to J_l, \; T)   + \sum_{J_u}   \zeta_I(J   \to  J_u, \; T)\big]  \; N_J + \sum_{J_l}   \zeta_I(\; J  \leftarrow   \; J_l, \; T) \; N_{J_l}        \nonumber \\  &&
+ \sum_{J_u } \zeta_S (J \leftarrow   J_u, \; T) \;  N_{J_u}  
\end{eqnarray} 
Therefore, we recover   the usual SEE for the level populations. 
 \subsection{Numerical results } 
 It is    important   to note   that it  is very difficult and sometimes impossible to treat collision processes involving heavy  ions like  Ba {\sc ii}  by standard quantum chemistry methods, regardless of whether they are simple or complex. A semi-classical  theory underlying the calculation of the depolarization and  polarization transfer rates by isotropic collisions of ions with
neutral hydrogen is  detailed  and validated   in Derouich et al. (\cite{Derouich_04}).    They found that the percentage of error on  the semi-classical rates    with respect to the available  quantum chemistry rates is 4\% to 7\%  (see Sect. 6 of Derouich et al. \cite{Derouich_04}).
 For computing the  collisional  rates of Ba {\sc ii}, we proceed level by level  since it is not possible 
to use an unique value of the so-called Uns\"old energy  (Uns\"old  1927, Uns\"old 1955) as for neutral atoms.  Although  the fact that this theory is of semi-classical nature, the close coupling is taken into account. We apply  our collisional numerical code   in order to calculate   the collisional transition 
matrix by solving the Schr\"odinger equation for  the $p$ and $d$-sates of Ba {\sc ii}.  Then, we obtain the transition  probabilities in the tensorial irreducible basis. 
Afterwards, these propabilities are integrated  over impact parameters and   Maxwellian distribution of   relative velocities to obtain the 
depolarization    and the transfer rates.  We  perform  calculations varying the temperature 
to obtain the best  analytical  fit to the collisional  rates. 

\begin{enumerate} 
\item In the case of the level $^2D_{5/2}$, the effect of the collisions is given in the  tensorial representation   by:
 \begin{eqnarray} \label{eq_1}
\big[\frac{d \; \rho_0^{0} (J=5/2)}{dt}\big]_{\rm coll} & = &  
  -  R^0 (5/2, \; T) \times \rho_0^{0} (J=5/2) \nonumber \\  &   & + \Big{[}  C_I(k=0 \; J=5/2 \leftarrow k=0 \; J_l=3/2, \; T)  \Big{]} \times \rho_0^{0} (J=3/2)   
\end{eqnarray}
 \begin{eqnarray} \label{eq_1}
\big[\frac{d \; \rho_q^{2} (J=5/2)}{dt}\big]_{\rm coll} & = &  
  -  R^2 (5/2, \; T) \times \rho_q^{2} (J=5/2)    \\
&& +   \big[C_I(k  =2 \; J=5/2 \leftarrow k =2 \; J_l=3/2, \; T)\Big{]}  \; \times\rho_q^2 (J=3/2)   \nonumber
\end{eqnarray}
 According to our calculations:
  \begin{eqnarray}  
 R^0 (5/2, \; T) (\textrm{s}^{-1}) &=& 
 \sqrt{\frac{2}{3}} \;  C_S(k=0 \; J=5/2 \to k=0 \; J_l=3/2, \; T)   \nonumber \\ &=&
 \sqrt{\frac{2}{3}}   \times \frac{2}{3}   \times 2.88  \times   10^{-9} \; n_{\textrm {\scriptsize H}} \Big(\frac{T}{5000}\Big)^{0.464}   \exp\frac{E_{ J=5/2}-E_{ J=3/2}}{k_B T}   
 \end{eqnarray}
 \begin{eqnarray}  
  C_I(k=0 \; J=5/2 \leftarrow k=0 \; J_l=3/2, \; T)  (\textrm{s}^{-1})  =  2.88  \times   10^{-9} \; n_{\textrm {\scriptsize H}} \Big(\frac{T}{5000}\Big)^{0.464}      
 \end{eqnarray}
 \begin{eqnarray}  
 R^2 (5/2, \; T) (\textrm{s}^{-1}) &=& 3.44 \times 10^{-9} \; n_{\textrm {\scriptsize H}} \Big(\frac{T}{5000}\Big)^{0.40} 
       +R^0 (5/2, \; T)  (\textrm{s}^{-1})
       \end{eqnarray}
         \begin{eqnarray}  
  C_I(k=2 \; J=5/2 \leftarrow k=2 \; J_l=3/2, \; T) (\textrm{s}^{-1}) &=& 1.27 \times 10^{-9} \; n_{\textrm {\scriptsize H}} \Big(\frac{T}{5000}\Big)^{0.26}    
\end{eqnarray}
 where  $T$ and  $n_{\textrm {\scriptsize H}}$ are respectively expressed in  Kelvins and cm$^{-3}$   and   $k_B$ is the Boltzmann constant.  Given the low degree of radiation anisotropy in the solar atmosphere, one can safely solve the SEE neglecting the elements with $k>2$, i.e. $\rho_q^{4} (J=5/2)$.
\item
In the case of the level $^2D_{3/2}$, the effect of the collisions is given in the  tensorial representation   by:
\begin{eqnarray} \label{eq_1}
\big[\frac{d \; \rho_0^{0} (J=3/2)}{dt}\big]_{\rm coll} & = &    -  R^0 (3/2, \; T) \times \rho_0^{0} (J=3/2) \nonumber \\ &&+ 
C_S(k  =0 \; J=3/2 \leftarrow k =0 \; J_u=5/2, \; T)  \; \times\rho_0^0 (J=5/2)   
\end{eqnarray}
\begin{eqnarray} \label{eq_1}
\big[\frac{d \; \rho_q^{2} (J=3/2)}{dt}\big]_{\rm coll} & = &    -  R^2 (3/2, \; T) \times \rho_q^{2} (J=3/2)  
 \nonumber \\
&& +  C_S(k  =2 \; J=3/2 \leftarrow k =2 \; J_u=5/2, \; T)  \; \times\rho_q^2 (J=5/2)    
\end{eqnarray}
where,   
\begin{eqnarray}  
  R^0 (3/2, \; T) (\textrm{s}^{-1}) & = &  \sqrt{\frac{3}{2}}  \times 2.88 \times 10^{-9} \; n_{\textrm {\scriptsize H}} \Big(\frac{T}{5000}\Big)^{0.464}   \\
     C_S(k  =0 \; J=3/2 \leftarrow k =0 \; J_u=5/2, \; T)  (\textrm{s}^{-1})   &=&  \\  \frac{2}{3}   &\times& 2.88  \times    10^{-9}  n_{\textrm {\scriptsize H}} \Big(\frac{T}{5000}\Big)^{0.464}   \exp\frac{E_{ J=5/2}-E_{ J=3/2}}{k_B T}  \nonumber \\
  R^2 (3/2, \; T) (\textrm{s}^{-1})  =  2.53  &\times&10^{-9} \; n_{\textrm {\scriptsize H}} \Big(\frac{T}{5000}\Big)^{0.41} + R^0 (3/2, \; T) (\textrm{s}^{-1}) \\
C_S(k  =2 \; J=3/2 \leftarrow k =2 \; J_u=5/2, \; T) (\textrm{s}^{-1})&=&  \\  \frac{2}{3}  \times 1.27 &\times& 10^{-9}  n_{\textrm {\scriptsize H}} \Big(\frac{T}{5000}\Big)^{0.26}  \times \exp\frac{E_{ J=5/2}-E_{ J=3/2}}{k_B T}  \nonumber
\end{eqnarray}
\item
In the case of the level $^2P_{3/2}$, one has:
\begin{eqnarray} \label{eq_1}
\big[\frac{d \; \rho_0^{0} (J=3/2)}{dt}\big]_{\rm coll} & = &    -  R^0 (3/2, \; T) \times \rho_0^{0} (J=3/2) \nonumber \\
&& +  \big[C_I(k  =0 \; J=3/2 \leftarrow k =0 \; J_l=1/2, \; T)\big]     \times \rho_0^{0} (J=1/2)
\end{eqnarray}
\begin{eqnarray} \label{eq_1}
\big[\frac{d \; \rho_q^{2} (J=3/2)}{dt}\big]_{\rm coll} & = &    -  R^2 (3/2, \; T) \times \rho_q^{2} (J=3/2),     
\end{eqnarray}
we find that:
\begin{eqnarray}  
R^0 (3/2, \; T)  (\textrm{s}^{-1})&=&   \sqrt{ \frac{1}{2}}  \times  \frac{1}{2}  \times 7.44 \times 10^{-9} \; n_{\textrm {\scriptsize H}} \Big(\frac{T}{5000}\Big)^{0.38}    \exp\frac{E_{ J=3/2}-E_{ J=1/2}}{k_B T}   
\end{eqnarray}
\begin{eqnarray}  
C_I(k  =0 \; J=3/2 \leftarrow k =0 \; J_l=1/2, \; T) (\textrm{s}^{-1})&=&   7.44 \times 10^{-9} \; n_{\textrm {\scriptsize H}} \Big(\frac{T}{5000}\Big)^{0.38}   
\end{eqnarray}
\begin{eqnarray}  
R^2 (3/2, \; T)  (\textrm{s}^{-1})&=&   6.82 \times 10^{-9} \; n_{\textrm {\scriptsize H}} \Big(\frac{T}{5000}\Big)^{0.40}    +   R^0 (3/2, \; T) (\textrm{s}^{-1})
\end{eqnarray}
\item
In the case of the level $^2P_{1/2}$, the effect of the collisions is:
\begin{eqnarray} \label{eq_1}
\big[\frac{d \; \rho_0^{0} (J=1/2)}{dt}\big]_{\rm coll} & = &    -  R^0 (1/2, \; T) \times \rho_0^{0} (J=1/2) \nonumber \\
&& +  \big[C_S(k  =0 \; J=1/2 \leftarrow k =0 \; J_u=3/2, \; T)\big]     \times \rho_0^{0} (J=3/2)     
\end{eqnarray}
where:
\begin{eqnarray}  
R^0 (1/2, \; T)  (\textrm{s}^{-1})&=&   \sqrt{ \frac{1}{2}}  \times   7.44 \times 10^{-9} \; n_{\textrm {\scriptsize H}} \Big(\frac{T}{5000}\Big)^{0.38}     
\end{eqnarray}
\begin{eqnarray}  
C_S(k  =0 \; J=1/2 \leftarrow k =0 \; J_u=3/2, \; T)  (\textrm{s}^{-1})&=&   \\   \frac{1}{2}  \times 7.44  &\times& 10^{-9} \; n_{\textrm {\scriptsize H}} \Big(\frac{T}{5000}\Big)^{0.38}    \exp\frac{E_{ J=3/2}-E_{ J=1/2}}{k_B T} \nonumber
\end{eqnarray}
\item
Finally, elastic collisions with neutral hydrogen do not affect the  ground  level $^2S_{1/2}$:\footnote{In linear   polarization studies,   singlet  levels with total angular momentum $J$= 1/2 are immune to   collisions   involving transitions inside the electronic state $^2S_{1/2}$. But if   the 18\% odd isotopes of Ba {\sc ii}   are not neglected, the   levels $J$ = 1/2 are split into hyperfine levels due to coupling with  nuclear spin $I$ = 3/2:  hyperfine levels $F$ = 1 and $F$ = 2 can be aligned and can be affected by collisions. }
\begin{eqnarray} \label{eq_1}
\big[\frac{d \; \rho_0^{0} (J=1/2)}{dt}\big]_{\rm coll} & = &  0     
\end{eqnarray}
\end{enumerate}
\section{Implication in the Hanle effect diagnostics}  
   First of all, we mention that, since we are not solving the radiative transfer problem, our results have to be considered as   complementary informations to the models     taking   into account radiative transfer  without    collisions  and realistic multilevel models. We introduce the collisional effect together with radiative rates and we determine the  emergent fractional polarization  in the limit of
tangential observation in a plane-parallel atmosphere where the   cosine of the heliocentric
angle $\mu$ $\simeq$ 0.

\begin{figure}
\begin{center}
\includegraphics[width=8cm]{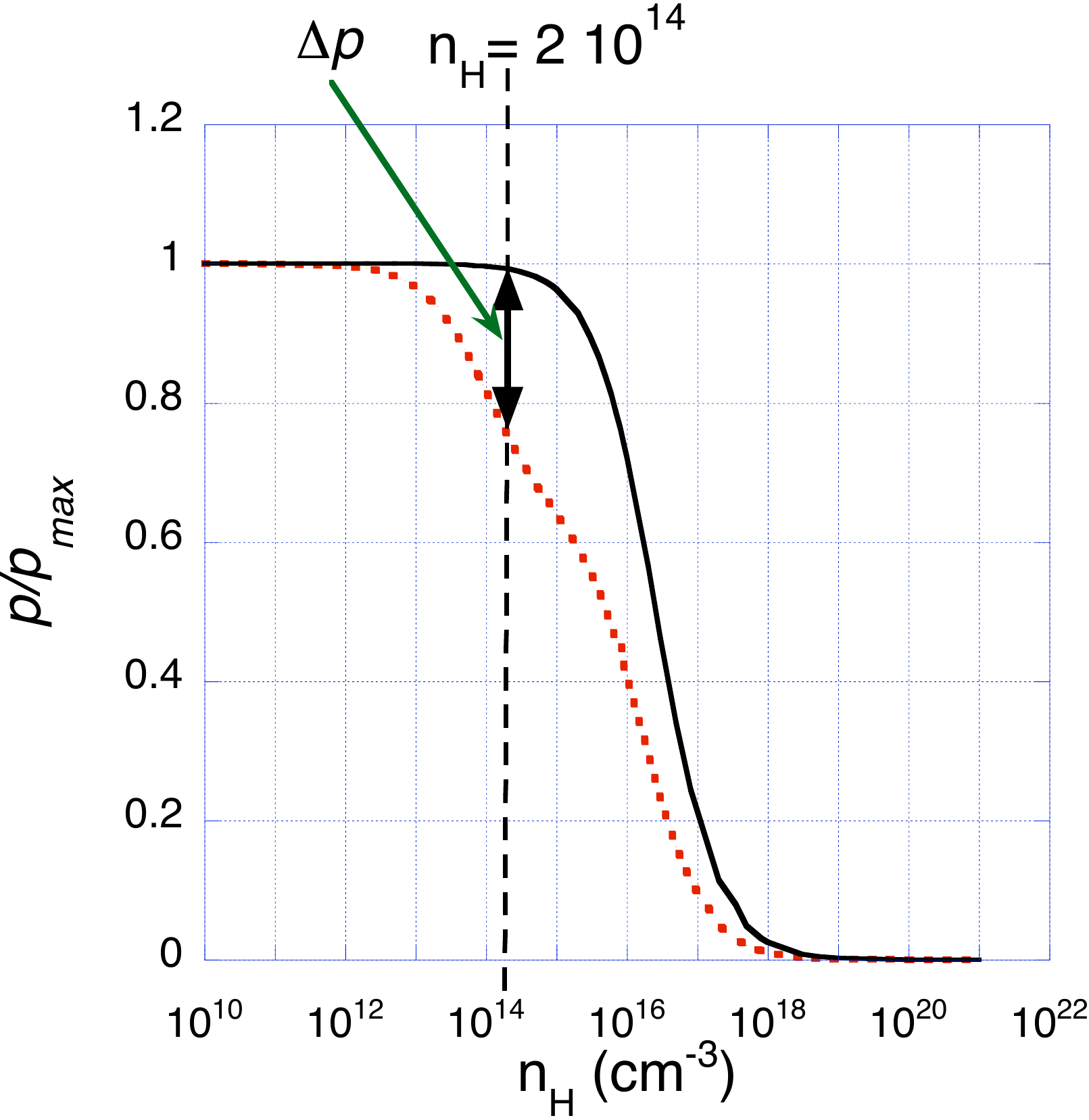}   
\end{center}
\caption[]{Emergent   linear polarization amplitudes $p$ divided by the zero-collisions polarization $p_{max}$  versus the density of neutral hydrogen  $n_{\textrm {\scriptsize H}}$. 
(1) Full line: variation of the linear polarization in the framework of the simplified model. (2) Dotted line: same as in (1) but in the framework of the 5 levels-5 lines model.}
\label{PoverPmax}
\end{figure}

If   we calculate the polarization in the framework of the 5 levels-5 lines model as a function of the neutral hydrogen density n$_\textrm{\scriptsize{H}}$ (see Fig. \ref{PoverPmax}), one could remark that collisions start to play a notable role for   $n_{\textrm {\scriptsize H}} \sim 3 \times  10^{13}$ cm$^{-3}$.  The alignment of the metastable levels $^2D_{3/2}$ and $^2D_{5/2}$ start to diminish due to  collisions and thus the polarization of the  \ion{Ba}{ii} line at ${\lambda}4554$ decreases.  Indeed, 
   for $n_{\textrm {\scriptsize H}} \sim 3 \times  10^{13}$ cm$^{-3}$, the ratio of the  polarization degree $p$ divided by the zero-collisions polarization $p_{max}$    is  $\sim$ 0.9 (i.e. a collisional depolarization of $\sim$ 10\%). 
  However, in the framework of the simplified model (Fig. \ref{figure1}), a collisional depolarization of $\sim$ 10\% of the ${\lambda}4554$ line is attempted only where $n_{\textrm {\scriptsize H}} \sim 3 \times  10^{15}$ cm$^{-3}$ (i.e. $\sim$ 100 times larger). This is because the upper level $^2P_{3/2}$ of the ${\lambda}4554$  line start to be affected by collisions solely for densities n$_\textrm{\scriptsize{H}}> 10^{15}$ cm$^{-3}$ (see Fig. \ref{PoverPmax}).

An estimate height of formation of the  ${\lambda}4554$  line  is $h \sim 800$ km which corresponds to a neutral hydrogen density $n_{\textrm {\scriptsize H}} \sim 2 \times  10^{14}$ cm$^{-3}$ and a temperature of the formation  of the lines $T  \sim$ 5350K (e.g. model C of Vernazza et al. 1981).   In these typical conditions of formation of the \ion{Ba}{ii} line at ${\lambda}4554$,  the polarization degree calculated using the simplified atomic model of Fig. \ref{figure1}  is  practically insensitive to collisions. This may yield, incorrectly, to the conclusion that collisions do not affect the Ba {\sc ii} line at ${\lambda}4554$. In the contrary, we find that, using  the 5 levels-5 lines atomic model, the polarization  degree is decreased by   $\sim$ 25\% because of the   collisions.     
Therefore, the collisions are an important ingredient to quantitatively interpret this line.  This is the main conclusion of the present work.

 The polarization in the absence of collisions is called $p_{max}$.  Since the impact approximation is well satisfied for  collisions between neutral hydrogen atoms  and perturbed ions in the solar atmosphere, the collisional  rates are simply  proportional to the hydrogen density (see their analytical expressions in Sect. 3).  To study the effect of collisions on the polarization degree, we    report in Fig. \ref{PoverPmax}  the  variation of the ratio $p/p_{max}$ as a function of the   hydrogen density. In Fig. \ref{PoverPmax}, we make together the results obtained for  the simplified model (full line) and for a more realistic model  (dotted line). 
We define the  percentage of error   in evaluating   the collisional sensitivity of the ${\lambda}4554$  line as 
 \begin{eqnarray}  
\Delta p=  \Big(1-({p/p_{max}})_\textrm{5 levels-5 lines model}\Big)   \times 100   
\end{eqnarray}   
In Fig. \ref{PoverPmax} we present $\Delta p$ for $n_{\textrm {\scriptsize H}} \sim 2 \times  10^{14}$ cm$^{-3}$.

 To assess the sensitivity of the microturbulent magnetic field determination to the collisions,    we  proceed as follow:
  \begin{enumerate}
 \item In the absence of collisions, we determine   the  polarization degree $p_0$ which corresponds to the critical magnetic field of the  \ion{Ba}{ii} ${\lambda}4554$ line, $B_0 \simeq 9$ G.
\footnote{
$B_0$ is the   magnetic field strength
 for which one may expect a sizable change
of the scattering polarization signal with respect to the
unmagnetized reference case.} 
 \item 
 In the absence of collisions, for each value of  an eventual collisional inaccuracy $\Delta p$ determined in Eq. (28), the magnetic field $B$ giving the polarization $p=p_0 \times (1+\Delta p/100)$ is retained.  In other words, we introduce a  perturbation $\Delta p$ to the polarization  $p_0$ and we calculate   the corresponding  magnetic field. We notice that, in general, the effect of the collisions  can be overestimated or neglected implying that  the polarization should be written generally as $p=p_0 \times (1\pm \Delta p/100)$. Here, we investigate only the more typical  case where the collisions are neglected which corresponds to $p=p_0 \times (1+\Delta p/100)$.
  \item The percentage of error   in evaluating the magnetic field   due to the neglecting of the collisions is defined by:
  \begin{eqnarray}  
\Delta B=  \Big|1-(B/B_{0})\Big|   \times 100   
\end{eqnarray}  
We obtain $\Delta B$ for each value of $\Delta p$ (i.e. for  each $n_{\textrm {\scriptsize H}}$). $\Delta B$ reaches the asymptotic value $100\%$ if   $B=2B_0$ or larger. 
 \end{enumerate}  
The choice  of the couple ($B_0$, $p_0$) as  a   starting point  before perturbing the polarization degree by a $\Delta p$ is arbitrary. An other starting couple  should give a rather similar $\Delta B$   provided that the magnetic field is suitable in the sense that it is well included in the  domain of sensitivity of the \ion{Ba}{ii} ${\lambda}4554$ line to the Hanle effect.     
   
\begin{figure}
\begin{center}
\includegraphics[width=8cm]{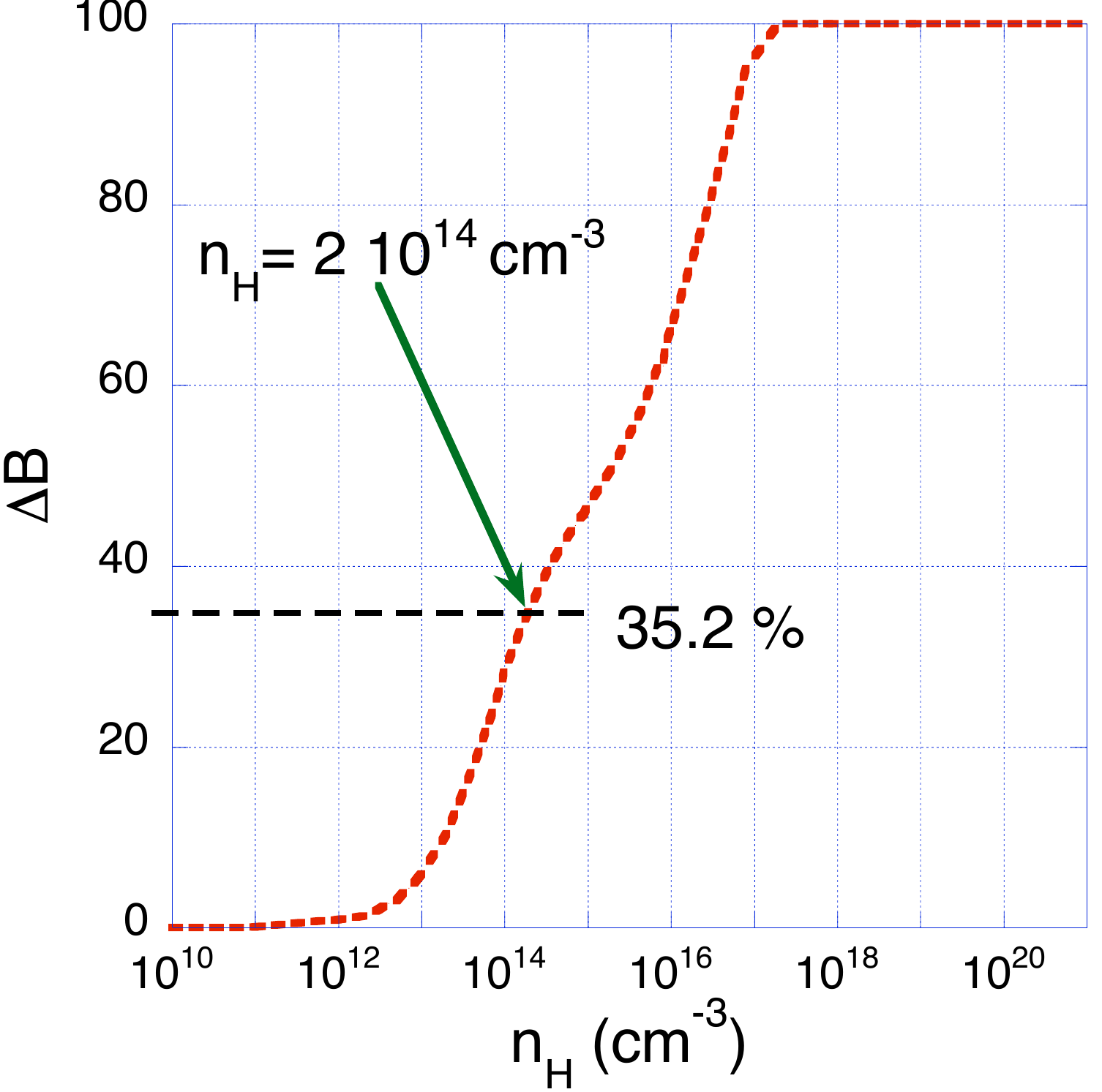}   
\end{center}
\caption[]{Percentage of   error on the magnetic field determination due to the neglecting of  the collisions with neutral hydrogen as a function of $n_{\textrm {\scriptsize H}}$.}
\label{ErrorMagnPercent}
\end{figure}

  In Fig.  \ref{ErrorMagnPercent}, we  show  the     percentage of error  $\Delta B$  as a function of $n_{\textrm {\scriptsize H}}$. For instance, at $n_{\textrm {\scriptsize H}} \sim 2 \times  10^{14}$ cm$^{-3}$, where the neglecting of collisions induces   an  overestimation of  the polarization    by  $\sim$ 25\%, the value of the magnetic field is    overestimated by $\sim$ 35\%.   As it could be easily seen in Fig.  \ref{ErrorMagnPercent},   depolarizing collisions can be neglected for n$_{\textrm {\scriptsize H}} <  \sim    10^{12}$  cm$^{-3}$ since the percentage error on the determination of the magnetic field  is almost zero. At  the asymptotic values of n$_{\textrm {\scriptsize H}}$  (i.e. n$_{\textrm {\scriptsize H}}$ $>\sim    10^{17}$  cm$^{-3}$), the calculation of the  scattering polarization is highly affected by  the neglecting of the  collisions  and the information about the magnetic field is completely lost (the  percentage of error is  $\sim$ 100\%).

\section{Concluding comments}
 It is well known that collisions with neutral hydrogen are   essential   to model    the  processes governing the formation of  many polarized lines in the solar photosphere   (D$_1$ and D$_2$  lines of \ion{Na}{i}; Ti {\sc i}  ${\lambda}4536$;   \ion{Sr}{i}  ${\lambda}4607$; etc). We show in this work
that, although  the neutral hydrogen is  about ten times less abundant  in the low chromosphere  than in the photosphere, the polarization  of the chromospheric  ${\lambda}4554$ line  decreases significantly  due to collisions with hydrogen atoms. 
 
 Conclusions about the role of the collisions should not be  only based 
   on a simple comparison of the inverse lifetime of the  upper level of the transition  and  the $D^2$  coefficients.
    Although this can give useful indications,   realistic conclusion   should be inferred only from a full introduction of the depolarization and   collisional transfer         rates  in the SEE for multi-level models.

Lines like    \ion{Ba}{ii}  ${\lambda}4554$ having  upper $np$-level   energy larger than the energy of the ($n$-1)$d$-level cannot be treated with a simplified model approximation ($n$ is the principal quantum number). For instance,  to model quantitatively  the K \ion{Sr}{ii}  ${\lambda}4078$ line  one should take into account the alignment of the $d$-sates and the   infrared triplet \ion{Sr}{ii} ${\lambda}10036$, \ion{Sr}{ii} ${\lambda}10327$, and \ion{Sr}{ii} ${\lambda}10914$. We notice that the K \ion{Sr}{ii}  ${\lambda}4078$ line  was examined by Bianda et al. (1998),  but  they  
used the traditional Van der Waals approach to calculate collisional rates and neglected the collisional depolarization of the long lived $d$-level.
 Interestingly, although    the  \ion{Mg}{ii}  is  in the same isoelectronic sequence as  \ion{Ca}{ii}, \ion{Sr}{ii}, and \ion{Ba}{ii}, the energy of the upper $p$-level of the K  \ion{Mg}{ii}  ${\lambda}2796$ line  is smaller than the energy of the $d$-level. This is fortunate because  the modeling of this line could be safely performed using a simplified atomic model neglecting the  role of the $d$-level.  Furthermore, the  K  \ion{Mg}{ii}  ${\lambda}2796$ line is expected to be strongly polarized  since the curve showing the anisotropy factor as a function of ${\lambda}$ reaches its maximum around ${\lambda} \sim 2800$ $\AA$  (see Fig. 2   of Manso Sainz \& Landi Degl'Innocenti  2002). In addition, in  given physical conditions, collisional effects  are clearly smaller for \ion{Mg}{ii}  than for \ion{Sr}{ii} and \ion{Ba}{ii}. Observations of this line are  difficult  from the ground-based telescopes but it could be observed with the help of high sensitivity modern polarimeters attached to space missions.

To point out trends in the depth dependence of the magnetic field, Derouich et al. (2006) interpreted 
{\it spatially-resolved} observations of the  photospheric \ion{Sr}{i}  ${\lambda}4607$ line. In order to extend their diagnostic    over larger parts of the solar atmosphere,   quantitative interpretation  of  the \ion{Ba}{ii}  ${\lambda}4554$ line,  also observed with spatial resolution, would be highly interesting.  To quantitatively study this line, one has to account for  partial frequency redistribution   effects (e.g. Uitenbroek  \& Bruls 1992 and Rutten  \& Milkey 1979). Here also collisions should play an important role since, besides their   effects on the atomic polarization, they  could change the frequency of the \ion{Ba}{ii} photons.  It remains a challenge to develop a general theory for partial frequency redistribution of polarized radiation in the presence of arbitrary magnetic fields and  including  the effects  of   collisions in a multilevel picture  with/without hyperfine structure.  

\begin{acknowledgements}
I would like to thank Andr\'es Asensio Ramos and Javier Trujillo Bueno
 for provinding numerical code used for the  SEE resolution.  An anonymous
referee  is thanked for critical comments that improved the presentation of this work.
\end{acknowledgements}


\end{document}